\documentclass[useAMS,usenatbib,twocolumn]{mn2e} \usepackage{psfig}
\usepackage{natbib} \usepackage{epsf} \usepackage{graphicx}

\bibpunct{(}{)}{;}{a}{}{,}
\begin{document}

\def\eps{\varepsilon} \def\aap{A\&A} \def\apj{ApJ} \def\apjl{ApJL}
\def\mnras{MNRAS} \def\aj{AJ} \def\nat{Nature} \def\aaps{A\&A Supp.}
\def\me{m_\e}
\def\lesssim{\mathrel{\hbox{\rlap{\hbox{\lower4pt\hbox{$\sim$}}}\hbox{$<$}}}}
\def\gtrsim{\mathrel{\hbox{\rlap{\hbox{\lower4pt\hbox{$\sim$}}}\hbox{$>$}}}}

\newcommand{\hq}{\hbar} \newcommand{\epsB}{\varepsilon_{\rm B}}
\newcommand{\epsCMB}{\varepsilon_{\rm cmb}}

\title[Structure Formation in Inhomogeneous Dark Energy Models]
{Structure Formation in Inhomogeneous Dark Energy Models}
\author[N. J. Nunes and D. F. Mota]{N. J. Nunes$^1$\thanks{
E-mail: n.j.nunes@qmul.ac.uk} and D. F. Mota$^{2}$\thanks{ E-mail:
mota@astro.ox.ac.uk}\\ $^1$ School of Mathematical Sciences, Queen
Mary, University of London, Mile End Road, London E1 4NS, U.K.\\
$^2$Astrophysics, Department of Physics, University of Oxford, Keble
Road, Oxford, OX1 3RH, U.K.}

\date{\today }

\maketitle
\begin{abstract}
We investigate how inhomogeneous quintessence models may have a
specific signature even in the linear regime of large scale structure
formation.  The dynamics of the collapse of a dark matter halo is
governed by the value or the dynamical evolution 
of the dark energy equation of state, the energy density's 
initial conditions and its
homogeneity nature in the highly non--linear regime. These have a
direct impact on the redshift of collapse, altering in consequence the
linearly extrapolated density threshold above which structures will
end up collapsing.  We compute this quantity for minimally
coupled and coupled quintessence models, examining two extreme
scenarios: first, when the quintessence field does not
exhibit fluctuations on cluster scales and below -- homogeneous dark energy;
and second, when the field inside the overdensity collapses
along with the dark matter -- inhomogeneous dark energy.  One shows that
inhomogeneous dark energy models present distinct features which may
be used to confront them with observational data, for instance galaxy
number counting.  
Fitting formulae for the linearly extrapolated density threshold 
above which structures will end up collapsing are provided for 
models of dark energy with constant equation of state.
\end{abstract}
\pagerange{\pageref{firstpage}--\pageref{lastpage}} \pubyear{2004}
\label{firstpage}

\section{Introduction}\label{sec:intro}

Measurements of the luminosity-redshift 
relationship from observations of supernovae of type Ia (SNIa)   \citep{riess,perl,knop}, 
the matter power spectrum of large scale structure as inferred 
from galaxy redshift surveys like the Sloan Digital Sky Survey (SDSS)   \citep{tegmark} 
and the 2dF Galaxy Redshift Survey (2dFGRS)   \citep{colless}, 
and the anisotropies  in the Cosmic Microwave Background Radiation (CMBR)   \citep{spergel}, 
reveal that a mysterious constituent with negative pressure,
so--called dark energy, accounts for seventy percent of today's
mass--energy budget and is causing the expansion of the universe to
accelerate.

Despite of its major importance in explaining the astrophysical data, 
the nature of dark energy is one of the greatest mysteries of modern cosmology. 
The simplest candidates for this unknown entity include the
cosmological constant (see e.g. \cite{carroll}) and a scalar
field, commonly designated quintessence, which varies across space and
changes with time (see e.g.  \cite{ratra}).

Though the nature of dark energy is unknown, 
one can still try to infer its properties
from its effects on cosmic structure formation.
In fact, the behaviour of linear perturbations in a scalar field and its effect
on large structure formation has been investigated by a number of authors
(see e.g. \citep{ferreira}).
However, the behaviour of quintessence during the non--linear
gravitational collapse is not well understood and is currently under
investigation (see e.g. \cite{wetterich02,amendola,mota,maor,peng,percival}).  
Usually, it is assumed that the quintessence field does not
exhibit density fluctuations on cluster scales and below. The reason
for this assumption is that, according to linear perturbation theory,
the mass of the field is very small (the associated wavelength of the
particle is of the order of the Hubble radius) and, hence, it does not
feel matter overdensities of the size of tenth of a Mpc or smaller
\citep{wang}.

The assumption of neglecting the effects of matter perturbations on
the evolution of dark energy at small scales is indeed a good
approximation when perturbations in the metric are very small.
However, care must be taken when extrapolating the small--scale
linear--regime results to the highly non--linear regime.  Then,
locally the flat FRW metric is no longer a good approximation to
describe the geometry of overdense regions.  Highly non--linear matter
perturbations could, in principle, modify the evolution of
perturbations in dark energy considerably, and these could, in turn,
back-react and affect the evolution of matter overdensities.  Moreover,
it is natural to think that once a dark matter overdensity decouples
from the background expansion and collapses, the field inside the
cluster feels the gravitational potential inside the overdensity and
its evolution will be different from the background evolution.  This
is indeed a general feature of not only cosmological scalar fields
whose properties depend on the local density of the region they ``live
in'' \citep{mota1,justin1,mota2,justin2,clifton}, but also of massless
particles, such as, photons (example of astrophysical effects are:
Lensing, Riess-Schiama effect, Sachs-Wolfe effect).

\cite{bean} suggested that the quintessence field could have an
important impact in the highly non---linear
regime. \cite{wetterich01,wetterich02,arbey} noted that the
quintessence field could indeed be important on galactic scales. It
was put forward by \cite{guz,guzman} that it could in fact be
responsible for the observed flat rotation curves in galaxies. Other
authors \citep{pad1,pad2,bag,causse} discussed more exotic models,
based on tachyon fields, and argued that the equation of state is
scale--dependent.

If it turns out that the effects of dark matter density perturbations
and metric influence perturbations of quintessence on small scales,
this could significantly change our understanding of structure
formation on galactic and cluster scales.

\cite{mota} have shown that properties of halos, such as the density
contrast and the virial radius, depend critically on the form of the
potential, the initial conditions of the field, the time evolution of
its equation of state and on the behaviour of quintessence in highly
non--linear regions.  In reality, the dependence on the inhomogeneity
of dark energy is only important for some dark energy candidates. If
the dark energy equation of state $w$ is constant, the differences between
the homogeneous and inhomogeneous cases are small, as long as the
equation of state does not differ largely from $w=-1$
\citep{mota,maor}.

If dark energy is indeed a non-negligible component inside matter
overdensities, then the dynamics of structure formation may be
strongly dependent on the nature of dark energy.  Different dark
energy models will contribute in very particular ways to the
gravitational potential of the overdensity via the Poisson equation.
This may result in specific signatures in the formation of large scale
structures.

In this paper, we investigate how inhomogeneous quintessence models
may have a specific signature even in the linear regime of large scale
structure formation. In particular, we investigate how the time of
collapse is affected by the inhomogeneity of dark energy and how this
is reflected on the linearly extrapolated density threshold above
which structures will end up collapsing, i.e.  $\delta_c(z) =
\delta_L(z = z_{\rm col})$.  This work extends upon previous studies
in that we examine the evolution of matter overdensities as a function
of a time--varying dark energy equation of state and its homogeneity
nature in the non-linear regime.

The article is organised as follows. In section 2 we investigate
minimally coupled dark energy models and describe briefly the
spherical collapse model and its dependence on the homogeneity nature
of dark energy. In section 3 we generalise our analysis to the case of
coupled quintessence models.  A summary of our principal results is
given in section \ref{sec:conclusions}.

\section{Minimally Coupled Dark Energy Models}
We will consider a spatially flat Friedmann-Robertson-Walker Universe
with scale factor $a(t)$. The cosmic dynamics is determined by a
background pressureless fluid (corresponding to dark and visible
matter), radiation and dark energy. The governing equations of motion
are
\begin{eqnarray}
\dot{H} &=&- \frac{\kappa^2}{2} \left( \rho_B^{}+p_B^{} + \rho_{\phi}
+ p_{\phi} \right) \,, \\
\label{eqcontinuity}
\dot{\rho}_B^{} &=& -3 H (\rho_B^{}+p_B^{}) \,, \\
\label{rhophieq}
\dot{\rho}_{\phi} &=& -3H(\rho_{\phi}+ p_{\phi}) \,,
\end{eqnarray}
subject to the Friedmann constraint
\begin{eqnarray}
H^2 = \frac{\kappa^2}{3} \left(\rho_B^{} + \rho_{\phi}^{} \right) \,.
\end{eqnarray}
where $H$ is the ratio of expansion of the Universe $H = \dot{a}/a$,
and $\kappa^2 = 8 \pi G$.  $\rho_B^{}$ and $p_B^{}$ are, respectively, the
energy density and pressure of the background fluid (dust and
radiation).  In this work we consider two possible scenarios for the
nature of dark energy.  If the dark energy is a perfect fluid then its
energy density and pressure are related by the equation of state
$p_{\phi} = w_{\phi} \rho_{\phi}$ and $\rho_{\phi} =
\Omega_{\phi0}\rho_0/a^{3(w_{\phi}+1)}$. Alternatively, dark energy
can be described by a dynamical evolving scalar field rolling down its
potential $V(\phi)$. In this case, it is defined energy density and
pressure of a scalar field as, $\rho_{\phi}^{} = \dot{\phi}^2/2 + V(\phi)$
and $p_{\phi}^{} = \dot{\phi}^2/2 - V(\phi)$, respectively.  The equation of
motion for the scalar field is,
\begin{equation}
\label{eqscalar}
\ddot{\phi} = - 3 H \dot{\phi} - \frac{d V}{d \phi} \,.
\end{equation}
%

\subsection{Dark energy models} \label{sec:demod}
In this work we will be exploring dark energy models for which
$w_{\phi} = -1$ (the cosmological constant), $w_{\phi} = -0.8$,
$w_{\phi} = -1.2$ (phantom energy, \cite{caldwell}) and two cases
where the dark energy is the result of a slowly evolving scalar field
in a potential with two exponential terms (2EXP) (\cite{copel})
\begin{equation}
V(\phi) = V_0 \left( e^{\alpha \kappa \phi} + e^{\beta \kappa \phi}
\right) \,.
\end{equation}
We have chosen the pair $(\alpha,\beta) = (6.2,0.1)$ (a) and
$(\alpha,\beta) = (20.1,0.5)$ (b) as they both provide an
equation of state at present $w_{\phi0} = -0.95$ though having
distinct evolutions at higher redshift. The equation of state for
(a) approaches zero at a higher rate than for (b).  What
mainly distinguishes these two models is the contribution of the field
at high redshift. As can be seen from Table.~\ref{tabela1}, (a)
provides a contribution of dark energy that is non negligible at high
redshifts. We expect, therefore, that this model will provide features
with a stronger departure from a pure cosmological constant than the
remaining models.
\begin{table}
\begin{tabular}{|c c|ccc|ccc|}
\hline $\alpha$ & \bf $\beta$ & ~ & $w_{\rm eff}$ &~ & ~&
$\Omega_{\phi}$ &~ \\ & & $z = 0$ & $z=2$ & $z=5$ & $z = 0$ & $z=2$ &
$z=5$ \\ \hline 6.2 & 0.1 & --0.95 & --0.85 & --0.76 & 0.7 & 0.15 &
0.1 \\ 20.1 & 0.5 &--0.95 & --0.95 & --0.94 & 0.7 & 0.1 & 0.02 \\
\hline
\end{tabular}
\caption{\label{tabela1} Evolution of the effective equation of state
  and contribution of the dark energy energy density with redshift for
  two pairs of parameters of the 2EXP model.}
\end{table}

One should also emphasize that the values of the present value of the
equation of state and its running ($w_0$ and $w_0' = dw/dz(z=0)$) of
these models are within the current bounds determined from supernovae
Ia (SNIa) observations (\cite{riess2}). Models (a), $w = -0.8$ and $w = -1.2$ 
correspond to limiting cases consistent with these observations. We 
have chosen them to estimate the largest range of departures on 
$\delta_c$ from the cosmological constant case. See however, \cite{jassal} 
where it seems that model (a) is not consistent with WMAP data.

In Table \ref{tabela1}, we
show how the effective equation of state (i.e. weighted average of the
equation of state between redshift 0 and $z$) depends on redshift for
these models. We are assuming that for all of these models
$\Omega_{\phi 0} = 0.7$.

Other models of dark energy such as the inverse power law
(\cite{Zlatev:1998tr}) or SUGRA models (\cite{brax}) have negligible
contribution of the scalar field at high redshift, hence, they are not
entirely distinct from a $w_{\phi} = {\rm constant}$ dark energy and
we will not consider them here.

\subsection{Spherical collapse model}
In the remaining of this section we will be using the spherical 
collapse model to describe
the gravitational collapse of an overdense region in minimally
coupled dark energy models. The radius of the overdense region $r$ and
density contrast $\delta$ are related in this case by 
$1+ \delta = \rho_{{\rm m}c}/\rho_{\rm m} =
(a/r)^3$, where $\rho_{{\rm m}c}$ and $\rho_{\rm m}$ 
are the energy densities of
pressureless matter in the cluster and in the background,
respectively.  In the next section we will study the spherical
collapse model for coupled quintessence models in which case this
relation is modified.

The equation of motion for the non-linear evolution of
the density contrast is
\begin{eqnarray}
\label{deltaeq}
\ddot{\delta} &=& - 2 \frac{\dot{a}}{a} 
\dot{\delta} + \frac{\kappa^2}{2}\rho_{\rm
m}(1+\delta)\delta + \frac{4}{3} \frac{\dot{\delta}^2}{1+\delta}
\nonumber \\ &~& + \frac{\kappa^2}{2}\left[
(1+3w_{\phi_c})\rho_{\phi_c} - (1+3w_{\phi})\rho_{\phi} \right]
(1+\delta) \,,
\end{eqnarray}
where we have considered the possibility that dark energy also
clusters, i.e. $\rho_{\phi c} \neq \rho_{\phi}$. In general, the
evolution of $\rho_{\phi c}$ in the cluster can be written as
(\cite{mota})
\begin{equation}
\label{rhophiceq}
\dot{\rho}_{\phi_c} = -3 \frac{\dot{r}}{r}(\rho_{\phi_c}+ p_{\phi_c}) +
\Gamma_{\phi} \,
\end{equation}
where $\Gamma_{\phi}$ describes the dark energy loss of energy inside
the cluster.  Note that $\dot{r}/r$ is given by
\begin{equation}
\label{dotdelta}
\frac{\dot{r}}{r} = \frac{\dot{a}}{a} - 
\frac{1}{3}~\frac{\dot{\delta}}{1+\delta} \,,
\end{equation}
hence, the system of equations closes.

Following \citep{mota}, we study the two extreme limits for the
evolution of dark energy in the overdensity region. In the first we
assume that dark energy is homogeneous, i.e. the value of
$\rho_{\phi}$ inside the cluster is the same as in the background,
with
\begin{equation}
\label{Gammaeq}
\Gamma_{\phi} = -3 \left( \frac{\dot{a}}{a} - \frac{\dot{r}}{r}
 \right) (\rho_{\phi_c}+ p_{\phi_c}) \,,
\end{equation}
in which case Eqs.~(\ref{rhophieq}) and (\ref{rhophiceq}) are
equivalent. In the second limit, dark energy is inhomogeneous,
collapses with dark matter such that $\Gamma_{\phi} = 0$. In this
case, the equation of motion for the scalar field inside the cluster
is
\begin{equation}
\label{eqscalarc}
\ddot{\phi}_c = - 3 \frac{\dot{r}}{r} \dot{\phi_c} - \frac{d
V(\phi_c)}{d \phi_c} \,.
\end{equation}
Another form of $\Gamma_{\phi}$, with  a different limit and situation, was also considered in \citep{maor,peng}, but we will not consider it here.

The linear regime of Eq.~(\ref{deltaeq}) defines the linear density
contrast $\delta_L$ determined by the equation
\begin{eqnarray}
\label{deltaleq}
\ddot{\delta}_L = - 2 H \dot{\delta}_L
+ \frac{\kappa^2}{2}\left[ \rho_{m} \delta_L + (1+3w_{\phi}) \rho_{\phi} \, \delta_\phi+3 \rho_\phi \delta w_\phi \right]  \,, \nonumber \\
\end{eqnarray}
where we have defined $\delta_{\phi} = \delta \rho_{\phi}/\rho_{\phi}$.
In the homogeneous case, $\delta_\phi = \delta w_{\phi} = 0$. On the other, for inhomogeneous dark energy with a constant equation of state, $\delta w_{\phi} = 0$ and 
$\delta_\phi$ satisfies the equation of motion
\begin{equation}
\ddot{\delta}_\phi = - 2 H \dot{\delta}_\phi +
\frac{\kappa^2}{2}(1+w_\phi) \left[ \rho_{m} \delta_L +
 (1+3w_{\phi}) \rho_\phi \delta_\phi \right]  \,.
\end{equation}
When dark energy is sourced by a scalar field, we have 
\begin{eqnarray}
\delta \rho_\phi &=& \dot{\phi} \, \delta \dot{\phi} + \frac{dV}{d\phi} \, \delta \phi \,, \\
\delta w_\phi &=& (1-w_\phi) \left( -\frac{1}{V} \, \frac{dV}{d\phi} \, \delta \phi + \delta_\phi \right) \,,
\end{eqnarray}
where the perturbation of the field, $\delta \phi$, satisfies
\begin{equation}
\label{deltaphieq}
\delta \ddot{\phi} = -3 H \delta \dot{\phi} - \frac{d^2V}{d\phi^2} \, \delta \phi + \dot{\delta}_L \, \dot{\phi} \,.
\end{equation}
Equations (\ref{deltaleq}) and (\ref{deltaphieq}) are equivalent to the ones found in \cite{hwang}.

\begin{figure}
\includegraphics[width=8.8cm]{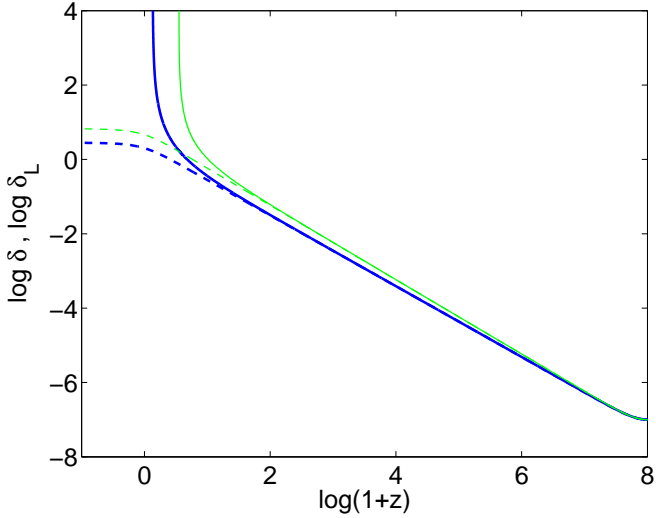}
\caption{Evolution of the density contrast $\delta$, and linear
  density contrast $\delta_L$ for the 2EXP model with 
$(\alpha,\beta)  = (6.2,0.1)$. Solid line: density contrast in
  homogeneous case; dashed line: linear density contrast in homogeneous
  case; thin solid line: density contrast in
  inhomogeneous case; thin dashed line: linear density contrast 
  in inhomogeneous
  case.
\label{fig:deltas}}
\end{figure}

In the case where the equation of state of dark energy is a constant,
one can derive that when $\Gamma_{\phi} = 0$ the non-linear evolution of the energy density of dark energy inside the collapsing region relates to the the one in the
background through
\begin{equation}
\label{rhophic}
\rho_{\phi_c} = \left(\frac{1+\delta}{1+\delta_i}\right)^{w_{\phi}+1}
\rho_{\phi} \,,
\end{equation}
for some initial perturbation $\delta_i$. 
In Fig.~\ref{fig:deltas} we compare the dependence of the 
density contrast and linear density contrast with 
redshift for a homogeneous and inhomogeneous scalar field dark energy
model. We see that in this case the density contrast and the
linear density contrast evolve with different slopes
from the homogeneous to the inhomogeneous cases. This is a
result of a non negligible contribution of the scalar field at high
redshifts in this particular model. In the case of a dark energy model
of constant equation of state, the various evolutions are essentially
indistinguishable except for the later stages when the dark energy
starts to dominate at low redshift. In general terms,
differences between homogeneous and inhomogeneous models
depend on the equation of state of dark energy (its value and if it is 
dynamical or not) and on the contribution of dark energy
for the total energy budget of the Universe at high redshifts.

We can now compute the linearly extrapolated density threshold above which
structures will end up collapsing, i.e.  $\delta_c(z) = \delta_L(z =
z_{\rm col})$. This quantity, is essential to compute the number of
collapsed structures following the Press-Schechter formalism
(\cite{press}). In our analysis we integrate the equations of motion
from very high redshifts ($1+z_i = 10^8$)
to minimize the effect of the decaying mode
on the evolution of the perturbations.

\begin{figure}
\includegraphics[width=8.8cm]{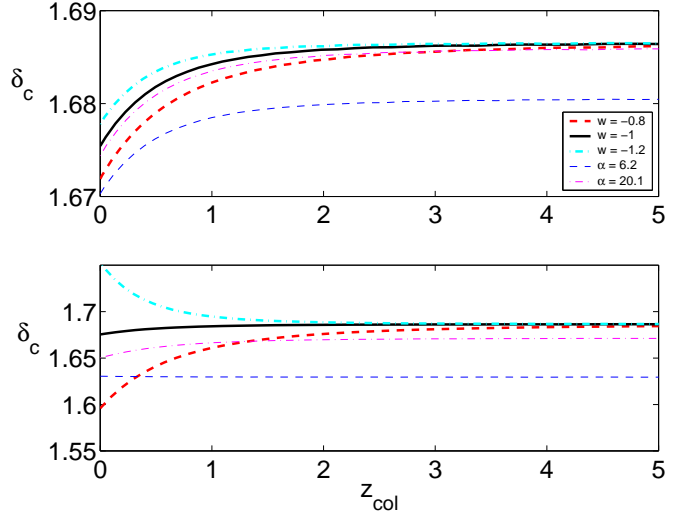}
\caption{Linear density contrast at the time of collapse in the
homogeneous (top panel) and inhomogeneous (bottom panel) dark energy
scenarios.
\label{fig:deltac}}
\end{figure}

For homogeneous scalar field dark energy,
the solution of Eq.~(\ref{deltaleq}) for a component $\phi$ of
fixed contribution to the total energy density  
$\Omega_{\phi} = 1- \Omega_{\rm m} = 3/\alpha^2$ (as it is the case of dark
energy models with an exponential term dominant at early times) and
neglecting the radiation component, is:
\begin{equation}
\label{deltacexp}
\delta_c(z_{\rm col}) = \frac{\delta_i}{2} \, (1+A^{-1})
\left(\frac{1+z_i}{1+z_{\rm col}}\right)^{(A-1)/4} \,,
\end{equation}
where $A = \sqrt{25-72/\alpha^2}$ and we have neglected the decaying
mode.  Given that $72/\alpha^2 = 24(1-\Omega_{\rm m})$ and that 
$1+z_{\rm} \propto t^{-2/3}$, we obtain the time dependence $\delta \propto t^m$, where
$m = (\sqrt{1+24\Omega_{\rm m}}-1)/6$, found by \cite{bag}, however, in the context of a scaling tachyon\footnote{We thank the anonymous referee for pointing this out to us.}. In an Einstein-de Sitter Universe (i.e. $\alpha = \infty$),
solving Eq.~(\ref{deltaeq}) with $1+z_i = 10^{8}$ for an initial
perturbation $\delta_i = 10^{-7}$, the overdensity region collapses at
redshift $z_{\rm col} = 2.558$, hence $\delta_c = 1.686$ (see
e.g. \cite{Padmanabhan}). This classical value was used in the seminal
paper of \cite{press}. For a model with one pure exponential potential
with $\alpha = 6.2$ we would obtain for the same initial conditions
$z_{\rm col} = 0.522$ and $\delta_c = 1.681$. This is the asymptotic
value of $\delta_c$ for the dark energy model considered,
$(\alpha,\beta) = (6.2,0.1)$, at high redshifts
(see upper panel of figure \ref{fig:deltac}).  At low redshifts,
when the dark energy component becomes important, the value of
$\delta_c$ decreases with the redshift of collapse.  

\begin{figure}
\includegraphics[width=8.8cm]{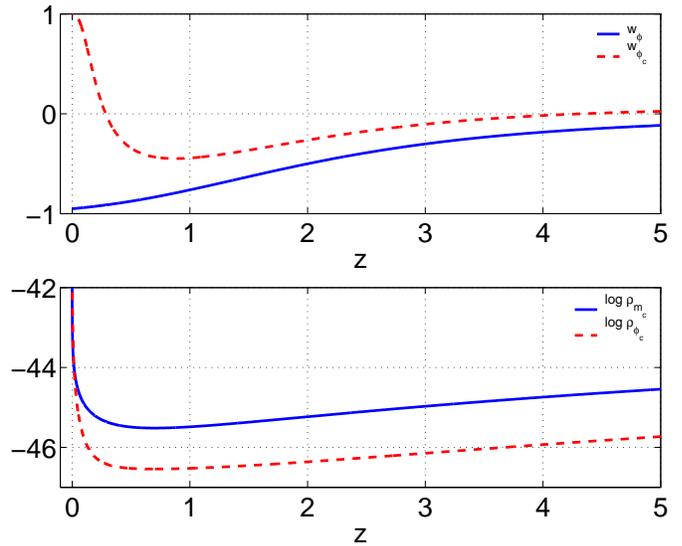}
\caption{\label{wphi} Upper panel: evolution of the 
dark energy equation of state
inside the overdense region (dashed line) and in the background 
(solid line) for a cluster that collapses today ($z_{\rm col} = 0)$.
This figure results
from numerically integrating the equations of motion of the scalar
field with the 2EXP model with $(\alpha,\beta) = (6.2,0.1)$. 
The turn around occurs at 
$z_{\rm ta} = 0.7$ which is shortly before the redshift at which the
equation of state inside the overdensity itself turns its evolution 
towards unity in opposition to its counterpart in the background that
continues to decrease to more negative values. Lower panel: evolution of the dark energy (dashed line) and dark matter (solid line) energy densities in the overdense region. The dark energy contribution only becomes dominant very close to the collapse when $w_{\phi_c}$ approaches unity.
}
\end{figure}

The computation of the value of $\delta_c$ at the time of collapse for 
inhomogeneous dark energy must 
consider in addition the terms in $\delta_\phi$ and $\delta w_{\phi}$ in Eq.~(\ref{deltaleq}).
We can set them to vanish as initial conditions, however, they are going to evolve in cosmological times, in fact, for a pure exponential potential, $\delta_L$ and $\delta_\phi$ will approach an attractor solution. Indeed, defining the dimensionless quantities:
\begin{eqnarray}
\psi \equiv \frac{\kappa \delta \phi}{\delta_L} \,, \hspace{0.5cm}
\sigma \equiv \frac{\kappa \delta \phi'}{\delta_L} \,, \hspace{0.5cm}
\tau \equiv \frac{\delta_L'}{\delta_L} \,,
\end{eqnarray}
where a prime means differentiation with respect to $\ln a$,
we can rewrite Eqs.(\ref{deltaleq}) and (\ref{deltaphieq}) in the form of a system of first order differential equations
\begin{eqnarray}
\psi' &=& \sigma - \tau \psi \,, \\
\sigma' &=& -\frac{9}{2}\psi - \frac{3}{2} \sigma + \frac{3}{\alpha} \tau - \tau \sigma \,, \\
\tau' &=& \frac{9}{2\alpha} \psi + \frac{6}{\alpha} \sigma -\frac{1}{2} \tau -\tau^2 + \frac{3}{2}\left(1-\frac{3}{\alpha}\right) \,.
\end{eqnarray}
This system has stable critical points at 
\begin{eqnarray}
\psi_c  = \frac{3}{7\alpha} \,, \hspace{0.5cm}
\sigma_c  = \frac{3}{7\alpha} \,, \hspace{0.5cm}
\tau_c = 1 \,,
\end{eqnarray}
where we have used the well known scaling solution result for an exponential potential $\kappa \phi' = 1/\alpha$ and $\kappa^2 V = 9H^2/2\alpha^2$ \citep{copeland}. Substituting back into $\delta_\phi$ and $\delta w_{\phi}$ we obtain that $\delta_\phi = -\delta_L/14$ and $\delta w_{\phi} = 5 \delta_L/14$ independent of $\alpha$. Upon substitution into 
Eq.~(\ref{deltaleq}) we find the interesting result
\begin{equation}
\ddot{\delta}_L = -2H\dot{\delta}_L + \frac{3}{2} H^2 \delta_L \,,
\end{equation}
i.e., the linear overdensity evolves in the attractor as if there is only dust in the Universe. However we should not expect to obtain $\delta_c = 1.686$ as in the Einstein-de Sitter Universe, for two reasons. Firstly, the system takes a finite amount of time before reaching the attractor solution. Secondly, the equation of state evolves inside 
the collapsing region from $w = 0$ at high redshifts to $w = 1$ at the
time of collapse (see Fig.~\ref{wphi}), thus increasing $\rho_\phi$ in the cluster 
favoring the clustering of dark matter. Hence, for the same initial conditions, 
the collapse occurs earlier, when the growth factor is smaller and 
therefore $\delta_c < 1.686$.

One can understand why the equation of state of dark energy evolves 
inside the overdensity region by inspecting equation (\ref{eqscalarc}). 
Indeed, after the turn around $\dot{r}/r$
becomes negative switching the frictional into an anti-frictional term
in the equation of motion 
of the scalar field. As the scalar potential becomes
less and less important the kinetic energy of the field approaches 
the asymptotic evolution $\dot{\phi}^2 \propto r^{-6}$ (therefore an
effective $w_{\phi_c} = 1$) which would in principle
overtake the energy density of pressureless matter as 
$\rho_{{\rm m}c} \propto r^{-3}$. In reality, this will never 
happen, since the
singularity in the collapse will not be reached.  The dark matter halo
will virialise much before and a dynamical equilibrium will 
be reached, where the halo has a constant radius (the virial radius).

It is important to notice that though the equation of state for dark 
energy inside the overdensity can become positive, it is still negative in 
the cosmological background (see figure \ref{wphi}). Hence, the  Universe's 
background expansion will not be affected
by the local behaviour of the dark energy inside clusters. The usual late 
time accelerated expansion still occurs as normally measured by SNIa. 
The only effects due to
the positivity of the dark energy equation of state  occur only inside the 
overdensities \citep{mota}. Moreover, within the models investigated, 
the dark energy contribution inside the cluster is always subdominant 
when compared to the one of dark matter and it only becomes dominant 
very near the collapse. We should not expect, therefore, any unusual 
effects on the virialisation process resulting from the positivity 
of the equation of state inside the cluster, at least for those models
with negligible background dark energy contribution at high
redshifts. Signatures in lensing and X-ray observations 
\citep{lopes,ana,sereno} 
may be noticed though, if dark energy provides a non negligible
background 
contribution at high redshifts.

In Fig.~\ref{fig:deltac}, we show the dependence of $\delta_c$ on the
redshift of collapse for homogeneous and inhomogeneous dark energy.
Note that for the cases of inhomogeneous scalar field dark energy, the value of
$\delta_c$ at high redshifts is considerably lower than for $w_{\phi}
= {\rm constant}$ cases. It is worth noting that for an inhomogeneous
dark energy, the evolution of $\delta_c$ with redshift is
significantly altered for the 'phantom energy', with $\delta_c$
decreasing from 1.753 to 1.686 for increasing redshift.  These
variations of the density threshold will obviously modify the
predicted numbers of collapsed objects.

The reason why one finds a difference in $\delta_c$ when comparing
homogeneous and inhomogeneous dark energy models, becomes clear when
one analyses the Poisson equation for the gravitational potential, $\Phi$,
inside the overdensity, 
\begin{equation}
\nabla^2 \Phi = \frac{\kappa^2}{2} (\rho_{\rm total}+3 p_{\rm total}) \,.
\label{poisson}
\end{equation}
In the case of homogeneous models 
\begin{eqnarray}
\nabla^2 \Phi &=& \frac{\kappa^2}{2} \left[\rho_{{\rm m}_c} +
  (1+3w_{\phi})\rho_{\phi_c} \right] \nonumber \\
&=& \frac{\kappa^2}{2} \left[(1+\delta)\rho_{{\rm m}} +
  (1+3w_{\phi})\rho_{\phi} \right] \,,
\label{poisson1}
\end{eqnarray}
as $\rho_{\phi_c} = \rho_{\phi}$. However for inhomogeneous
models, making use of equation (\ref{rhophic}) for 
a constant equation of state, we have 
\begin{equation}
\nabla^2 \Phi = \frac{\kappa^2}{2}\left[(1+\delta)\rho_{{\rm m}} +
  (1+3w_{\phi})\left(\frac{1+\delta}{1+\delta_i}\right)^{1+w_\phi}
\rho_{\phi} \right] \,.
\label{poisson2}
\end{equation}
Comparing equations (\ref{poisson1}) and (\ref{poisson2}) one notices that
in the inhomogeneous case the contribution of the scalar field energy
density becomes increasingly important as the density contrast grows. Since
$w_{\phi}$ is generally negative, 
the source term in equation (\ref{poisson})
is smaller than in the homogeneous case. Conversely, if $w_{\phi}$ is
more negative than $-1$  the contribution of the scalar field is less
and less important and the source term is larger than in the
homogeneous case.
In conclusion, inhomogeneous dark energy models will contribute to the
gravitational potential inside the overdensities in a different way to
homogeneous ones. These will clearly affect the whole dynamics of the
formation of structures.  In particular the late stages of the
non-linear collapse of overdensities, such as the time of collapse,
virial radius and matter density contrast \citep{mota}.  In the case
of dark energy candidates with a dynamical equation of state, such as
scalar fields, the effects will be even more interesting. In these
cases the source term in equation (\ref{poisson}) will vary in time
since $w_{\phi_c}$ will change during the formation of the collapsing
overdensity, resulting in a more complex evolution for the
gravitational potential. Recall that $w_{\phi_c}$ may even become
positive and approach $w_{\phi_c} = 1$ during the late phase 
of collapse.  Hence the dark energy pressure in
equation (\ref{poisson}) will source the gravitational potential
positively, in opposition to the usual homogeneous models where
$w_{\phi}$ is always negative. The effects will be stronger at low
redshifts when dark energy dominates the universe density in the
background universe.

For a constant equation of state $w$ the following expressions provide
an accurate fit to the evolution of the density contrast as shown in
Fig.~\ref{fig:deltac}:
\begin{equation}
\delta_c(z) = \frac{3}{20} (12 \pi)^{2/3} \left[ 1+ \alpha x + \beta
  x^2 \right] \,,
\end{equation}
where $x = \log(\Omega_m(z))$ and for homogeneous dark energy we have
\begin{eqnarray}
\alpha &=& 0.0061w^2 + 0.0327w + 0.0403 \,, \\
\beta &=&  -0.0163w^2   -0.0294w   -0.0118 \,,
\end{eqnarray}
whereas for inhomogeneous dark energy it reads
\begin{eqnarray}
\alpha &=& 0.1198w^2  + 0.6226w  +  0.5170 \,, \\ 
\beta &=& 0.2022w^2  + 0.2877w  + 0.0860 \,.
\end{eqnarray}
These expressions are valid for $-1.2 < w< -0.6$.

\section{Coupled quintessence}
Let us now look at the case when the scalar field has a coupling with
all or part of the dark matter (see for
e.g. \cite{amendola,domenico,domenico2}).  In these models,
inhomogeneities in the quintessence field may appear due to two main
reasons: The first, which is the same as in minimally coupled models,
is the change of the local geometry of the region where the
overdensity ``lives in'', which is just the general relativistic
effect of the spacetime deformation. The second is the dragging of the
scalar field by the dark matter particles caused by the coupling
between them. This second effect is similar to the one which occurs in
scalar-tensor theories and leads to inhomogeneities in the scalar
field \citep{clifton,mota1}.  Due to these causes, it is then 
natural to expect stronger
inhomogeneity effects in this kind of models than in the minimally
coupled one.

The background quantities and the ones who live inside the collapsing
region are given by
\begin{eqnarray}
\label{eq:cq1}
\rho_{\rm um} &=& \rho_0\Omega_{\rm um0}
\left(\frac{a_0}{a_i}\right)^3 \left(\frac{a_i}{a}\right)^3 \,, \\
\label{eq:cq2}
\rho_{\rm cDM} &=& \rho_0\Omega_{\rm cDM0}
\left(\frac{a_0}{a_i}\right)^3 \left(\frac{a_i}{a}\right)^3
e^{B(\phi)-B(\phi_0)} \,, \\ 
\label{eq:cq3}
\rho_{{\rm um}c} &=& (1+\delta_i)
\rho_0\Omega_{{\rm um}c0} \left(\frac{a_0}{a_i}\right)^3
\left(\frac{r_i}{r}\right)^3 \,, \\ 
\label{eq:cq4}
\rho_{{\rm cDM}c} &=& (1+\delta_i)
\rho_0\Omega_{{\rm cDM}c0} \left(\frac{a_0}{a_i}\right)^3
\left(\frac{r_i}{r}\right)^3 e^{B(\phi_c)-B(\phi_0)} \,,
\nonumber \\ 
\end{eqnarray}
where the subscripts ``um'' and ``cDM'' mean uncoupled matter and
coupled dark matter, respectively. Uncoupled matter corresponds to
both baryons and to uncoupled dark matter. The function $B(\phi)$
represents the coupling between dark energy and dark matter. 
In the model discussed by
\cite{holden} and \cite{amendola1}, $B(\phi) = -C\kappa \phi$, where
$C$ is a constant. But other forms for this function have also 
been suggested  (\cite{domenico0,mainini}).

The total energy densities
inside the cluster and the background are therefore, $\rho_{\rm m} =
\rho_{\rm um} + \rho_{\rm cDM}$ and $\rho_{c} = \rho_{{\rm um}c} +
\rho_{{\rm cDM}c}$ which evolve accordingly to
\begin{eqnarray}
\dot{\rho}_{\rm m} &=& -3 \frac{\dot{a}}{a} \rho_{\rm m} +
                     \frac{dB}{d\phi} \rho_{\rm cDM} \dot{\phi} \,, \\
                     \dot{\rho}_{c} &=& -3 \frac{\dot{r}}{r} \rho_{c}
                     + \frac{dB}{d\phi_c} \rho_{{\rm cDM}c}
                     \dot{\phi_c} \,.
\end{eqnarray}
If we take for initial conditions $\phi_c(a = a_i) = \phi(a=a_i)$ we
can define $r_i = a_i/(1+\delta_i)^{1/3}$ and then we can write
$1+\delta \equiv \rho_{c}/\rho_{\rm m}$, as
\begin{equation}
1+\delta = \left(\frac{a}{r}\right)^3 \frac{\Omega_{{\rm
cDM}c0}e^{B(\phi_c)-B(\phi_0)} +\Omega_{{\rm um}c0}}{\Omega_{{\rm
cDM}0}e^{B(\phi)-B(\phi_0)}+\Omega_{{\rm um}0}} \,.
\end{equation}
It will become useful to write the ratios $G = G(\phi) \equiv \rho_{\rm
cDM}/\rho_{\rm m}$ and $G_c = G(\phi_c) \equiv \rho_{{\rm cDM}c}/\rho_{\rm
m}(1+\delta)$ in the following form:
\begin{eqnarray}
G(\phi) = \frac{\Omega_{{\rm cDM}0}e^{B(\phi)-B(\phi_0)}}
{\Omega_{{\rm cDM}0}e^{B(\phi)-B(\phi_0)}+\Omega_{{\rm um}0}} \,.
\end{eqnarray}

The time derivative of the density contrast will now have a component
coming from the coupling $dB/d\phi$ in the equations above
\begin{equation}
\dot{\delta} = 3(1+\delta)
\left[\frac{\dot{a}}{a}-\frac{\dot{r}}{r}\right] + (1+\delta) F(\phi)
\,,
\end{equation}
(compare with equation (\ref{dotdelta}))
where have defined $F(\phi)$ as
\begin{eqnarray}
(1+\delta) F(\phi) &=& \frac{dB}{d\phi_c} \frac{\rho_{{\rm
cDM}c}}{\rho_{\rm m}} \dot{\phi}_c - \frac{dB}{d\phi} \frac{\rho_{{\rm
cDM}}}{\rho_{\rm m}} \dot{\phi} (1+\delta) \,, \\ &=& (1+\delta)
\left[\frac{dB}{d\phi_c}G_c \dot{\phi}_c - \frac{dB}{d\phi}
G \dot{\phi} \right]\,,
\end{eqnarray}
and it results that $\dot{F}(\phi)$ is
\begin{eqnarray}
\dot{F}(\phi) &=& G \left[\frac{d^2B}{d\phi_c^2}\dot{\phi}_c^2 +
  \left(\frac{dB}{d\phi_c}\right)^2\dot{\phi}_c^2(1-G_c)+
  \frac{dB}{d\phi_c} \ddot{\phi}_c \right] \nonumber \\ &-&
  G\left[\frac{d^2B}{d\phi^2}\dot{\phi}^2 +
  \left(\frac{dB}{d\phi}\right)^2\dot{\phi}^2(1-G)+
  \frac{dB}{d\phi} \ddot{\phi} \right] \,. \nonumber \\
\end{eqnarray}
The equations of motion for the evolution of the scalar field
inside and the background are in this case:
\begin{eqnarray}
\label{eq:phicq}
\ddot{\phi} &=& - 3 \frac{\dot{a}}{a} \dot{\phi} - \frac{dV}{d\phi}
-\frac{dB}{d\phi} \rho_{\rm cDM} \, \\ 
\ddot{\phi}_c &=& -3
\label{eq:cqc}
\frac{\dot{r}}{r} \dot{\phi}_c - \frac{dV}{d\phi_c}
-\frac{dB}{d\phi_c} \rho_{{\rm cDM}c} +
\frac{\Gamma_{\phi}}{\dot{\phi}_c} \,.
\end{eqnarray}

Using these equations we are now able to obtain the modified equation
for the non-linear evolution of the density contrast
\begin{eqnarray}
\ddot{\delta} &=& - 2 \frac{\dot{a}}{a} \left[\dot{\delta}-(1+\delta)F
\right] + \frac{\kappa^2}{2}\rho_{\rm m}(1+\delta)\delta + \frac{4}{3}
\frac{\dot{\delta}^2}{1+\delta} \nonumber \\ &~& +
\frac{\kappa^2}{2}\left[ (1+3w_{\phi_c})\rho_{\phi_c} -
(1+3w_{\phi})\rho_{\phi} \right] (1+\delta) \nonumber \\ &~&
-\frac{2}{3} F \dot{\delta} + \frac{1}{3} (1+\delta)F^2 +
(1+\delta)\dot{F} \,.
\end{eqnarray}
From linearizing the expression of the density contrast one gets
\begin{eqnarray}
\ddot{\delta}_L &=& - 2 H (\dot{\delta}_L-f) + \dot{f} \nonumber \\
&~& + \frac{\kappa^2}{2}\left[ \rho_{m} \delta_L + (1+3w_{\phi_c})\delta_\phi ~\rho_{\phi} +3 \rho_\phi \delta w_\phi \right]  \,.
\end{eqnarray}
where $\delta \phi$ is determined through the equation of motion
\begin{eqnarray}
\delta \ddot{\phi} &=& -3 H \, \delta \dot{\phi} 
-\frac{dB}{d\phi} \, G\, \rho_m  \delta_L + (\dot{\delta}_L-f) \dot{\phi} \nonumber \\
&~& - \left[ \frac{d^2V}{d\phi^2} +\left(\frac{dB}{d\phi}\right)^2\,G\,(1-G)\, \rho_m + \frac{d^2B}{d\phi^2} \, G \, \rho_m \right]\delta \phi  \,, \nonumber \\ 
\end{eqnarray}
and $f$ is the linearization of $F$:
\begin{equation}
f = G \left[ \frac{dB}{d\phi} \, \delta \dot{\phi} + \left(\frac{dB}{d\phi}\right)^2(1-G)\dot{\phi} \, \delta \phi + \frac{d^2B}{d\phi^2}\, \dot{\phi} \, \delta \phi \right] \,.
\end{equation}

\begin{figure}
\includegraphics[width=8.4cm]{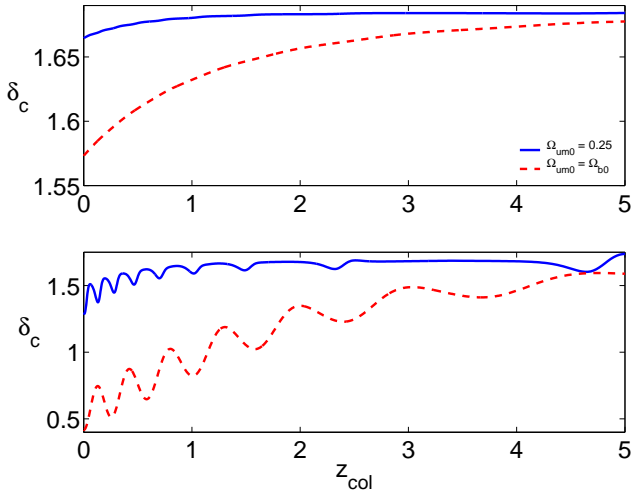}
\caption{Linear density contrast at the time of collapse in the
homogeneous (top panel) and inhomogeneous (bottom panel) coupled
quintessence scenarios. 
\label{fig:deltaccq}}
\end{figure}

Assuming a pure exponential
potential $V(\phi) = V_0 \exp(\alpha \kappa \phi)$ with $\alpha = 10$,
and $B(\phi) = -C\kappa \phi$, 
we compare in figure \ref{fig:deltaccq}
the evolution of the linear
density contrast at the time of collapse, $\delta_c$, between
homogeneous and inhomogeneous dark energy. We take into
consideration the case when all the dark matter is coupled with the
scalar field, or in other words, that the uncoupled matter is only the
baryons, hence $\Omega_{\rm um 0} = \Omega_{b0}$; and the possibility that
only a small fraction of the dark matter feels the field in which case
we have taken $\Omega_{\rm um 0} = 0.25$ in figure \ref{fig:deltaccq}.
Similarly to what happens in minimally coupled models, there are clear
differences in the evolution of $\delta_c$ between inhomogeneous and
homogeneous models. As expected, these differences are more distinct
in the case where the amount of coupled dark matter is larger
(dashed-lines in figure \ref{fig:deltaccq}). 

A particular feature is
the oscillating behaviour of $\delta_c$ that is amplified in the inhomogeneous
case. These oscillations are a result of the oscillations performed 
by the scalar field around the late time attractor solution. As we can 
see from Eqs.~(\ref{eq:cq2}) and (\ref{eq:cq4}), these oscillations 
induce a corresponding oscillation in the 
$\rho_{\rm cDM}$ and $\rho_{{\rm cDM}c}$ components (see Fig.~7 of 
\cite{Copeland:2003cv}). 

\begin{figure}
\includegraphics[width=8.4cm]{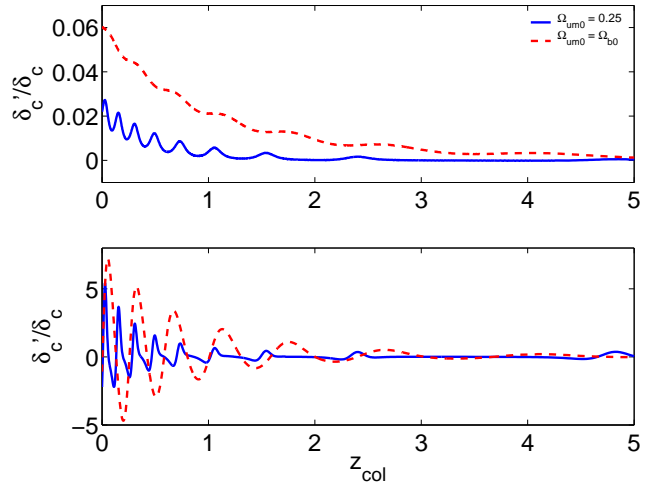}
\caption{Ratio $\delta_c'/\delta_c$ where prime is differentiation with 
respect to redshift for homogeneous (top panel) and inhomogeneous 
(bottom panel) coupled quintessence scenarios.
\label{fig:oscilcq}}
\end{figure}

Though these oscillations appear inexistent in the homogeneous scenario 
they are nonetheless present as can be seen by computing 
$\delta_c'/\delta_c = (d \delta_c/dz)/\delta_c$, see Fig.~\ref{fig:oscilcq}. 
They only seem inexistent because the mean $\delta_c$ evolves quicker 
than the amplitude of the oscillations. For the inhomogeneous case, however,
the scalar field enters a regime of amplified oscillations 
inside the overdensity after 
the turn around (as the $\dot{\phi}_c$ term in Eq.~(\ref{eq:cqc}) becomes
anti-frictional), hence its oscillations are progressively larger. 
Consequently,
the oscillations in $\delta_c$ are more probable of becoming visible.

\section{Summary and Conclusions}\label{sec:conclusions}

During the highly non-linear regime of cosmological perturbations,
when dark matter halos are formed, scalar fields may become
inhomogeneous.  Hence, evolving differently inside a collapsed
overdensity and in the background Universe.  The reason is simple:
Different local spacetime geometries lead to different local equations
of motion for the field. In the case of dark energy candidates, such
as quintessence, this may result on dark energy having a different
equation of motion in the background Universe and inside clusters. As
a consequence, the dynamics of non-linear structure formation in the
universe may show a distinct signature associated to the nature of
dark energy and a particular model \citep{mota}.

The behaviour of dark energy during the non-linear regime of structure
formation can be divided into two major groups: the inhomogeneous
models, which are the ones where the field inside the overdensity
behaves differently form the background Universe 
and the homogeneous ones, where dark energy is just a
smooth fluid which bath the whole Universe.

In this article we have investigated the effects of inhomogeneities in
dark energy on the time of collapse of a dark matter overdensity, and
how this is reflected on the linearly extrapolated density threshold
above which structures will end up collapsing, i.e.  $\delta_c(z) =
\delta_L(z = z_{\rm col})$.  We have studied both coupled and
minimally coupled scalar field candidates to dark energy as well as fluids
with a constant equation of state.  We have compared all these models
and we found distinct features among them when taking into account the
possible inhomogeneities in the dark energy fluid during the
non-linear regime.

In general  $\delta_c(z)$ depends on the particular model one considers. 
For constant equation of state candidates to dark energy there are minor 
differences between homogeneous and inhomogeneous models. The
exception to 
that are the 
phantom energy candidates, with $w_{\phi}<-1$. These 
present a quite distinct behaviour when one takes into account the 
inhomogeneities in the dark energy fluid.
However, it is on scalar fields candidates where inhomogeneities in
the dark energy fluid may result in more concrete features.
The later will depend on its  dynamical equation of state, which can
be scale dependent, and on the contribution of 
the field to the energy budget of the Universe both at low and high redshifts.
Coupled quintessence models behave similarly to the minimally coupled
field candidates, but naturally show  
a strong dependence on the amount of dark matter coupled to  the
scalar field and the strength of the coupling.

The reason for such a difference between inhomogeneous and homogeneous
models can be understood looking at the Poisson equation
(\ref{poisson}) which gives
the evolution of the gravitational potential inside the clusters.  In
opposition to the homogeneous models, 
one cannot neglect the
importance of dark energy inside the matter overdensities.  This leads
to an extra source term in the usual Poisson equation, which affects
the overall behaviour of the gravitational potential.  This effect is
specially important when the collapsing region enters into the
non-linear regime.  In particular, the time of collapse of the
overdensity depends on the dark energy model one considers and on the
equation of state inside it. These effects are reflected into linearly
extrapolated density threshold above which structures will end up
collapsing, $\delta_c$.

The results of this work may have important consequences on
cosmological studies of structure formation.  In particular one can
study its implications for the number density of dark matter halos,
their density profiles, galaxy number counting and dark matter halo
concentrations \citep{nelson,marc}.  These investigations are imperative for cosmological
studies that rely on these ingredients to measure dark
energy. Examples of these studies, for the case of homogeneous dark
energy models, include semi-analytical studies of strong lensing
statistics \citep{bart03,lopes} and weak lensing number counts
\citep{bart03}. And for inhomogeneous models include a study of dark
matter halo concentrations \citep{ana}.

Many other astrophysical phenomena
may reflect the inhomogeneities in dark energy fluid during the 
non-linear regime. 
For instance,  the merger of dark matter halos depend on the 
gravitational potential between them. 
Effects on the evolution and magnitude of the gravitational
potential,  may also affect peculiar velocities of galaxies inside
clusters.  
One should also point out here, that in low-density regions of the
universe, such as voids, inhomogeneous dark energy models, may play an
important role as well.  In those regions the local spacetime geometry
may be different from the usual Friedman-Robertson-Walker metric one
usually assumes for the background Universe.  It is natural to think
that dark energy inside voids may behave differently from its
background behaviour.  This may again affect the dynamics of
formation of voids in the universe. All these effects could in 
principle be detected using 
galaxy redshift surveys such as the Sloan Digital Sky Survey 
\citep{sdss,sdss1}.

\noindent\\

{\large \bf ACKNOWLEDGMENTS}\\
DFM is supported by Fundac\~{a}o Ci\^{e}ncia e a Tecnologia. NJN is
supported by the Particle Physics and Astronomy Research Council
(PPARC) and is thankful to  Nabila Aghanim and Ant\'{o}nio da Silva for
useful discussions and hospitality at IAS Orsay, during the initial stages
of this work. The authors also thank Ana Lopes for comments on the manuscript.

\bibliographystyle{apsrmp}

\begin{thebibliography}{99}

\bibitem[\protect\citeauthoryear{Abazajian et al.} {2003}]{sdss}
Abazajian K., {\it et al.}  [SDSS Collaboration], 2003,
Astron.\ J.\  126, 2081 

\bibitem[\protect\citeauthoryear{Alcubierre et al.} {2002}]{guzman}
Alcubierre A., Guzman F., Matos T., Nunez D., Urena-Lopez L.,
Wiederhold P., 2002, Class. Quant. Grav., 19, 5017.


\bibitem[\protect\citeauthoryear{Amendola} {2000}]{amendola1} Amendola
L., 2000, Phys.\ Rev.\ D 62 043511.

\bibitem[\protect\citeauthoryear{Amendola \& Tocchini-Valentini}
{2002}]{domenico0} Amendola L., and Tocchini-Valentini D., 2001, 
Phys.\ Rev.\ D 64 043509.

\bibitem[\protect\citeauthoryear{Amendola} {2003}]{amendola} Amendola
L., 2003, preprint astro-ph/0311175.

\bibitem[\protect\citeauthoryear{Amendola \& Tocchini-Valentini}
{2002}]{domenico2} Amendola L., and Tocchini-Valentini D., 2002,
Phys.\ Rev.\ D 66 043528.

\bibitem[\protect\citeauthoryear{Arbey et al.} {2001}]{arbey} Arbey A.,
Lesgourgues J., Salati P., 2001, Phys. Rev. D, 64, 123528.

\bibitem[\protect\citeauthoryear{Bagla et al.} {2003}]{bag} Bagla J.,
Jassal H., Padmanabhan T., 2003, Phys. Rev. D, 67, 063504.

\bibitem[\protect\citeauthoryear{Barreiro et al.} {2000}]{copel}
Barreiro T., Copeland E. $\&$ Nunes N.J., 2000, Phys. Rev. D, 61,
127301

\bibitem[\protect\citeauthoryear{Bartelmann et al.} {2003}]{bart03}
Bartelmann M., Meneghetti M., Perrotta F., Baccigalupi C., Moscardini
L., 2003, A$\&$A, 409, 449.

\bibitem[\protect\citeauthoryear{Bean $\&$ Magueijo} {2002}]{bean}
Bean R. $\&$ Magueijo J., 2002, Phys. Rev. D, 66, 063505.

\bibitem[\protect\citeauthoryear{Brax $\&$ Martin} {1999}]{brax} Brax
P. $\&$ Martin J. 1999, Phys. Lett. B, 468, 40.

\bibitem[\protect\citeauthoryear{Caldwell} {2002}]{caldwell} Caldwell
R., 2002, Phys.\ Lett.\ B 545 23.


\bibitem[\protect\citeauthoryear{Carroll et al.} {1992}]{carroll}
Carroll S.M., Press W.H. $\&$ Turner E.L., 1992,
Ann. Rev. Astron. Astrophys., 30, 499.

\bibitem[\protect\citeauthoryear{Causse} {2003}]{causse} Causse M.,
2003, preprint astro-ph/0312206.

\bibitem[\protect\citeauthoryear{Colless et al.}{1998}]{colless} Colless M., 1998, 
"First results from the 2dF galaxy redshift survey", astro-ph/9804079

\bibitem[\protect\citeauthoryear{Clifton, Mota $\&$ Barrow}
{2004}]{clifton} Clifton T., Mota D.F., Barrow J.D., 2004, preprint
gr-qc/0406001.

\bibitem[\protect\citeauthoryear{Copeland, Liddle $\&$ Wands}
{1998}]{copeland} 
Copeland E.J., Liddle A.R., Wands D., 1998,
Phys.\ Rev.\ D 57 4686.

\bibitem[\protect\citeauthoryear{Copeland et al.}{2004}]{Copeland:2003cv} Copeland E.J., Nunes N. J., and Pospelov M., 2004,
  Phys. Rev. D 69, 023501.



\bibitem[\protect\citeauthoryear{Ferreira $\&$ Joyce}
{1998}]{ferreira} Ferreira P, and Joyce M., 1998,
Phys.\ Rev.\ D 58  023503.

\bibitem[\protect\citeauthoryear{Guzman $\&$ Urena-Lopez} {2003}]{guz}
Guzman F., Urena-Lopez L. 2003, Phys. Rev. D, 68, 024023.


\bibitem[\protect\citeauthoryear{Holden $\&$ Wands} {2000}]{holden}
Holden D.J., and Wands D., Phys.\ Rev.\ D 61 043506.

\bibitem[\protect\citeauthoryear{Hwang $\&$ Noh} {2001}]{hwang}
Hwang J.~c., Noh H., Phys.\ Rev.\ D 64 103509.


\bibitem[\protect\citeauthoryear{Jassal et al.} {2005}]{jassal}
Jassal H., Bagla J. S., Padmanabhan T., 2005, Phys.Rev.D72:103503.

\bibitem[\protect\citeauthoryear{Khoury $\&$ Weltman} {2004}]{justin1}
Khoury J., Weltman A., 2004, Phys. Rev. D {69} 044026.

\bibitem[\protect\citeauthoryear{Khoury $\&$ Weltman} {2003}]{justin2}
Khoury J., Weltman A., 2003, preprint astro-ph/0309300.

\bibitem[\protect\citeauthoryear{Knop et al.} {2003}]{knop} Knop R.A.,
et al., ApJ, 2003, 598, 102.

\bibitem[\protect\citeauthoryear{Lopes $\&$ Miller} {2004}]{lopes}
Lopes A.M. $\&$ Miller L., 2004, MNRAS, 348, 519.

\bibitem[\protect\citeauthoryear{Lopes, Mota $\&$ Miller} {2004}]{ana}
Lopes A.M., Mota D. F. $\&$ Miller L., 2004, submitted to A$\&$A.


\bibitem[\protect\citeauthoryear{Maor $\&$ Lahav}{2004}]{maor}
Maor I., Lahav O., 2004, JCAP 0507:003.

\bibitem[\protect\citeauthoryear{Mainini $\&$ Bonometo}{2004}]{mainini}
Mainini R., Bonometo S.A.,
preprint astro-ph/0406114.


\bibitem[\protect\citeauthoryear{Manera \& Mota} {2004}]{marc} Manera M. $\&$ Mota
D.F. astro-ph/0504519.


\bibitem[\protect\citeauthoryear{Mota \& Barrow} {2004a}]{mota1} Mota
D.F., Barrow J.D., 2004a, Phys. Lett. B, 581, 141.

\bibitem[\protect\citeauthoryear{Mota $\&$ Barrow} {2004b}]{mota2}
Mota D.F., Barrow J.D., 2004b, MNRAS, 349, 281.

\bibitem[\protect\citeauthoryear{Mota $\&$ van de Bruck} {2004}]{mota}
Mota D.F. $\&$ van de Bruck C., 2004, A$\&$A, 421, 71.



\bibitem[\protect\citeauthoryear{Nunes, da Silva $\&$ Aghanim} {2005}]{nelson}
Nunes N. J., da Silva A. C. $\&$ Aghanim N., 2005, A$\&$A in print, 
e-Print Archive: astro-ph/0506043.



\bibitem[{Padmanabhan} ({1995})]{Padmanabhan} Padmanabhan, T. 1995,
{\it Structure formation in the universe}, Cambridge University Press

\bibitem[\protect\citeauthoryear{Padmanabhan} {2002}]{pad2}
Padmanabhan T., 2002, Phys. Rev. D 66, 021301.

\bibitem[\protect\citeauthoryear{Padmanabhan $\&$ Choudhury}
{2002}]{pad1} Padmanabhan T., Choudhury T., 2002, Phys. Rev. D, 66,
081301.


\bibitem[\protect\citeauthoryear{Wang} {2005}]{peng}
Wang P., 2005, astro-ph/0507195.


\bibitem[\protect\citeauthoryear{Percival} {2005}]{percival}
Percival W., 2005,astro-ph/0508156.

\bibitem[\protect\citeauthoryear{Perlmutter et al.} {1999}]{perl}
Perlmutter S., et al., 1999, ApJ, 517, 565.

\bibitem[\protect\citeauthoryear{Press $\&$ Schechter} {1974}]{press}
Press W., Schechter P., 1974, Astrophys.\ J.\ 187 425.

\bibitem[\protect\citeauthoryear{Ratra $\&$ Peebles} {1988}]{ratra}
Ratra B., Peebles P.,1988, ApJ, 325, L17.

\bibitem[\protect\citeauthoryear{Riess et al.} {2001}]{riess} Riess
A.G., et al., 2001, ApJ., 560, 49.

\bibitem[\protect\citeauthoryear{Riess et al.} {2004}]{riess2} Riess
A.G. {\it et al.}  [Supernova Search Team Collaboration],2004,
Astrophys.\ J.\ {607}, 665.

\bibitem[\protect\citeauthoryear{Sereno $\&$ Longo} {2004}]{sereno} 
Sereno M. $\&$ Longo G., 2004, MNRAS, 354, 1255.

\bibitem[\protect\citeauthoryear{Spergel et al.} {2003}]{spergel}
Spergel D.N., et al., 2003, ApJ Suppl., 148. 175.


\bibitem[\protect\citeauthoryear{Tegmark et al.}{2004}]{tegmark} Tegmark et al., 2004, Phys. Rev. D69, 103501.

\bibitem[\protect\citeauthoryear{Tocchini-Valentini \& Amendola }
{2002}]{domenico} Tocchini-Valentini D., and Amendola L., 2002, Phys.\
Rev.\ D {65} 063508

\bibitem[\protect\citeauthoryear{Wang $\&$ Steinhardt} {1998}]{wang}
Wang L.M. $\&$ Steinhardt P.J., 1998, ApJ, 508, 483.

\bibitem[\protect\citeauthoryear{Wetterich} {2001}]{wetterich01}
Wetterich C., 2001, Phys. Lett. B, 522, 5.

\bibitem[\protect\citeauthoryear{Wetterich} {2002}]{wetterich02}
Wetterich C., 2002, Phys. Rev. D, 65, 123512.

\bibitem[\protect\citeauthoryear{Zehavi et al.} {2002}]{sdss1}
Zehavi I., et al. [SDSS Collaboration], 2002,
Astrophys.\ J.\  571, 172. 

\bibitem[\protect\citeauthoryear{Zlatev et al.} {1999}]{Zlatev:1998tr}
Zlatev I., Wang L., and Steinhardt P., 1999, Phys.\ Rev.\ Lett.\ 82,
896.

\end{thebibliography}

\label{lastpage}
\end{document}